\documentclass[%
 reprint,
 amsmath,amssymb,
 aps,
]{revtex4-2}

\usepackage{graphicx}
\usepackage{dcolumn}
\usepackage{bm}
\usepackage{color}
\usepackage[colorlinks,linkcolor=blue,citecolor=blue]{hyperref}
\usepackage{bm}
\usepackage{amsmath}
\usepackage{float}

\begin{document}

\preprint{APS/123-QED}

\title{Dressed bound states at chiral exceptional points}

\author{Yuwei Lu}
\affiliation{School of Physics and Optoelectronic Engineering, Foshan University, Foshan 528000, China}
\affiliation{School of Physics and Optoelectronics, South China University of Technology, Guangzhou 510641, China}
\author{Haishu Tan}%
\email[Corresponding Author: ]{tanhaishu@fosu.edu.cn}
\affiliation{School of Physics and Optoelectronic Engineering, Foshan University, Foshan 528000, China}

\author{Zeyang Liao}
\email[Corresponding Author: ]{liaozy7@mail.sysu.edu.cn}
\affiliation{State Key Laboratory of Optoelectronic Materials and Technologies, School of Physics, Sun Yat-sen University, Guangzhou 510275, China}


\date{\today}

\begin{abstract}
Atom-photon dressed states are a basic concept of quantum optics. Here, we demonstrate that the non-Hermiticity of open cavity can be harnessed to form the dressed bound states (DBS) and identify two types of DBS, the vacancy-like DBS and Friedrich-Wintgen DBS, in a microring resonator operating at a chiral exceptional point. With the analytical DBS conditions, we show that the vacancy-like DBS occurs when an atom couples to the standing wave mode that is a node of photonic wave function, and thus is immune to the cavity dissipation and characterized by the null spectral density at cavity resonance. While the Friedrich-Wintgen DBS can be accessed by continuously tuning the system parameters, such as the atom-photon detuning, and evidenced by a vanishing Rabi peak in emission spectrum, an unusual feature in the strong-coupling anticrossing. We also demonstrate the quantum-optics applications of the proposed DBS. Our work exhibits the quantum states control through non-Hermiticity of open quantum system and presents a clear physical picture on DBS at chiral exceptional points, which holds great potential in building high-performance quantum devices for sensing, photon storage, and nonclassical light generation.
\end{abstract}

\maketitle


\section{Introduction}

Dressed states are a hallmark of strong atom-photon interaction \cite{qo}, which provide a basis for coherent control of quantum states and give rise to a rich variety of important technologies and applications, such as quantum sensing \cite{nlwl}, entanglement transport \cite{prbe,prbce}, photon blockade for quantum light generation \cite{valle,scpma,xky}, and many-body interaction for scalable quantum computing and quantum information processing \cite{np2015,prbcpr}. Dressed states with slow decay, i.e., narrow linewidth, are appealing in practical applications. Though the linewidth of dressed states is the average of atomic and photonic components, it often limited by the latter since the linewidth of quantum emitter (QE) is much smaller than the cavity at cryogenic environment. Therefore, a natural approach to reduce the linewidth of dressed states is by mean of high-$Q$ cavity, which, however, is often at the price of large mode volume \cite{opticahq,ol1998} or requires elaborate design \cite{prl2017,sa2018,acsp2016}. Furthermore, light trapping and release are time reversal processes in linear time-invariant systems, thus a cavity with high $Q$ in general leads to low excitation efficiency, which is undesirable in practical applications. These disadvantages stimulate the exploration of alternative scheme to suppress the decay of dressed states.

Despite leakage is inevitable for optical resonators, it also opens up new avenues for manipulating light-matter interaction by exploiting the non-Hermitian degeneracies \cite{science,npPT}, known as exceptional points. The presence of exceptional points renders the exotic features to the system dynamics due to the reduced dimensionality of underlying state space at exceptional points \cite{mrm,oe,epjd}. Particularly, previous studies have shown that the coalescence of counterclockwise (CCW) and clockwise (CW) modes in whispering-gallery-mode (WGM) microcavity gives rise to a special type of exceptional points, called chiral exceptional points (CEP) \cite{pnas,prl2014}, which exhibits the unprecedented degree of freedom in state control, such as the quantum and optical states with chirality \cite{mrm,hughesacsp,pnas} and the spontaneous emission enhancement associated with squared Lorentzian response \cite{oe,prr,prl2022,prb2017}. 
\begin{figure*}[t]
\centering\includegraphics[width=0.7\linewidth]{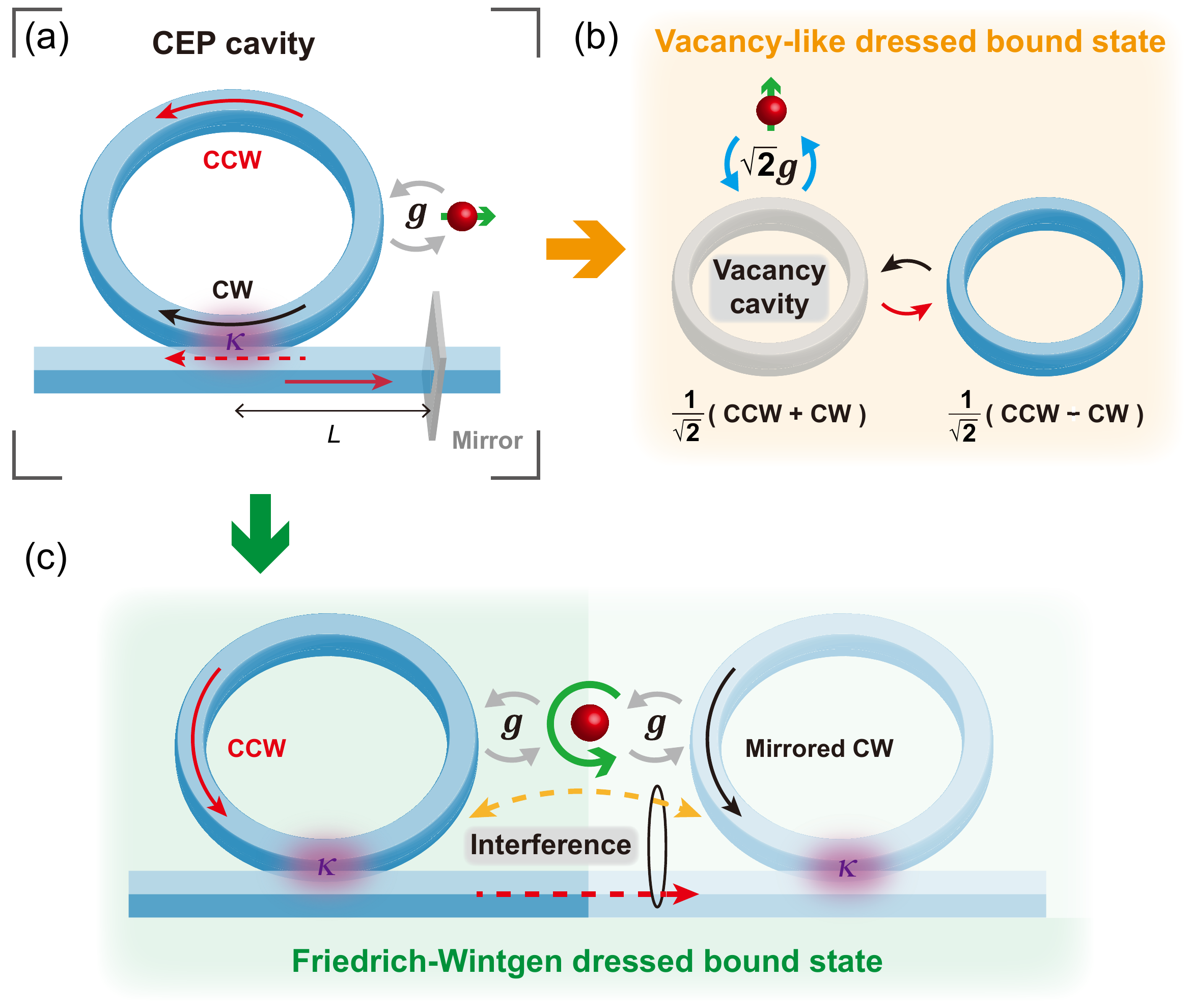}
\caption{(a) Schematic of the CEP cavity where a WGM microring coupled to a QE and a waveguide with a mirror at the right end. (b) Illustration of the origin of vacancy-like DBS: the standing wave mode that the QE couples to is a node of wavefunction. (c) Illustration of the formation of Friedrich-Wintgen DBS via the destructive interference between the coupling pathways mediated by the QE and the waveguide. The CW mode is flipped to another CCW mode via mirror symmetry. Accordingly, the linearly polarized QE becomes circularly polarized.}
\label{fig1}
\end{figure*}

In this work, we propose and identify the formation of dressed bound states (DBS) in an open microring resonator with CEP, which we call CEP cavity hereafter. A theoretical framework is established to unveil the origin and derive the analytical conditions of DBS. We show that DBS in CEP cavity can be classified into two types, the vacancy-like DBS \cite{vlbs} and Friedrich-Wintgen DBS \cite{scpma,prl2008,opticaes,bicnature}. The vacancy-like DBS has a unique feature that its condition is irrespective of atom-photon coupling strength since the cavity mode the atom coupled to is a node of photonic wavefunction. By contrast, DBS with Friedrich-Wintgen origin depends on the system parameters, such as the frequency detuning and coupling strength between different system components, which are required to fulfill the condition of destructive interference between two coupling pathways. We also discuss the characteristics of spontaneous emission (SE) spectrum and dynamics associated with DBS and demonstrate the corresponding quantum-optics applications.

\section{Results and Discussion}
\subsection{Model and Theory}
The CEP cavity we study is depicted in Fig. \ref{fig1}(a), where a WGM microring resonator is coupled to a semi-infinite waveguide with a perfect mirror (i.e., unity reflectivity) at the end. The mirror results in chiral coupling from CCW mode to CW mode and creates a CEP \cite{prr}. A linearly polarized QE couples to CEP cavity with coupling strength $g$. We assume that the QE is embedded inside the cavity, thus its coupling to free space via modes other than cavity modes is suppressed. The quantum dynamics of the cavity QED system is described by the extended cascaded quantum master equation (see Refs. \cite{qme,prrLD} and also Appendix. \ref{aa} for detailed derivation)

\begin{equation}\label{eq1}
\begin{gathered}
\frac{d}{dt} \rho=-i[H, \rho]+\kappa \mathcal{L}\left[c_{ccw}\right] \rho+\kappa \mathcal{L}\left[c_{cw}\right] \rho \\
+\kappa\left(e^{i \phi}\left[c_{ccw} \rho, c_{cw}^{\dagger}\right]+e^{-i \phi}\left[c_{cw}, \rho c_{ccw}^{\dagger}\right]\right)
\end{gathered}
\end{equation}
where $\mathcal{L}[O] \rho=O \rho O^{\dagger}-\left\{O^{\dagger} O, \rho\right\} / 2$ is the Liouvillian superoperator for dissipation of operator $O$. The Hamiltonian is given by $H=H_0+H_I$, where the free Hamiltonian $H_0$ and the interaction Hamiltonian $H_I$ read
\begin{equation}
H_0=\omega_0 \sigma_{+} \sigma_{-}+\omega_c c_{c c w}^{\dagger} c_{c c w}+\omega_c c_{c w}^{\dagger} c_{c w}
\end{equation}
\begin{equation}
H_I=g\left(c_{c c w}^{\dagger} \sigma_{-}+\sigma_{+} c_{c c w}\right)+g\left(c_{c w}^{\dagger} \sigma_{-}+\sigma_{+} c_{c w}\right)
\end{equation}
where $\sigma_{-}$ is the lowering operator of QE, while $c_{ccw}/c_{cw}$ is the bosonic annihilation operator for CCW/CW mode. $\omega_0$ and $\omega_c$ are the transition frequency of QE and the resonance frequency of cavity modes, respectively. Considering the high-$Q$ feature of WGM modes, the intrinsic decay of cavity is omitted, thus its dissipation is determined by the evanescent coupling $\kappa$ to the guided mode of waveguide. The second line of Eq. (\ref{eq1}) describes the chiral coupling that the CW mode is driven by the output field from the CCW mode, where $\phi=2 \beta L$ is the accumulated phase factor of light propagation, with $\beta$ and $L$ being the propagation constant of waveguide and the distance between the waveguide-resonator junction and the mirror, respectively.

We consider the SE process that there is at most one photon in the system and the resonant QE-cavity coupling ($\omega_0 = \omega_c$). The equations of motion in the single-excitation subspace can be obtained from Eq. (\ref{eq1})
\begin{equation}\label{eq2}
\frac{d}{d t} \vec{p}=-i \mathbf{M}_c \vec{p}
\end{equation}
with $\vec{p}=\left[\left\langle\sigma_{-}\right\rangle,\left\langle c_{c c w}\right\rangle,\left\langle c_{c w}\right\rangle\right]^T$ and the matrix $\mathbf{M}_c$

\begin{equation}\label{eq3}
\mathbf{M}_c=\left[\begin{array}{ccc}
\omega_c & g & g \\
g & \omega_c-i \frac{\kappa}{2} & 0 \\
g & -i \kappa e^{i \phi} & \omega_c-i \frac{\kappa}{2}
\end{array}\right]
\end{equation}

The emission spectrum is experimentally relevant and also critical to understand the quantum dynamics of a QE. Therefore, we investigate the spectrum properties of DBS via the SE spectrum of QE, which can be measured via fluorescence of QE and is defined as $S(\omega)=\lim _{t \rightarrow \infty} \operatorname{Re}\left[\int_0^{\infty} d \tau\left\langle\sigma_{+}(t+\tau) \sigma_{-}(t)\right\rangle e^{i \omega \tau}\right]$ \cite{qo,prb2012}, where $\left\langle\sigma_{+}(t+\tau) \sigma_{-}(t)\right\rangle$ can be calculated from the equations of single-time averages (Eqs. (\ref{eq2})-(\ref{eq3})) using the quantum regression theorem \cite{qo}
\begin{equation}
\frac{d}{d \tau}\left[\begin{array}{c}
\left\langle\sigma_{+}(\tau) \sigma_{-}(0)\right\rangle \\
\left\langle\sigma_{+}(\tau) c_{c c w}(0)\right\rangle \\
\left\langle\sigma_{+}(\tau) c_{c w}(0)\right\rangle
\end{array}\right]=-i\mathbf{M}_c\left[\begin{array}{c}
\left\langle\sigma_{+}(\tau) \sigma_{-}(0)\right\rangle \\
\left\langle\sigma_{+}(\tau) c_{c c w}(0)\right\rangle \\
\left\langle\sigma_{+}(\tau) c_{c w}(0)\right\rangle
\end{array}\right]
\end{equation}
The above equations can be solved via the Laplace transform with the initial conditions $\left\langle\sigma_{+}(0) \sigma_{-}(0)\right\rangle=1$, $\left\langle\sigma_{+}(0) c_{ccw}(0)\right\rangle=0$, and $\left\langle\sigma_{+}(0) c_{cw}(0)\right\rangle=0$. The SE spectrum of QE is expressed as (see Appendix. \ref{ab} for detailed derivation)
\begin{equation}
S(\omega)=\frac{1}{\pi} \frac{\Gamma(\omega)}{\left[\omega-\omega_c-\Delta(\omega)\right]^2+\left[\frac{\Gamma(\omega)}{2}\right]^2}
\end{equation}
where $\Gamma(\omega)=-2g^2\operatorname{Im}[\chi(\omega)]$ is the local coupling strength and $\Delta(\omega)=g^2\operatorname{Re}[\chi(\omega)]$ denotes the photonic Lamb shift, with $\chi(\omega)$ being the response function of CEP cavity
\begin{equation}\label{chi}
\chi(\omega)=\frac{2}{\left(\omega-\omega_c\right)+i \frac{\kappa}{2}}-\frac{i \kappa e^{i \phi}}{\left[\left(\omega-\omega_c\right)+i \frac{\kappa}{2}\right]^2}
\end{equation}
The SE dynamics of QE can be retrieved from $\mathcal{F}[S(\omega)]$, the Fourier transform of SE spectrum.

Eqs. (\ref{eq1})-(\ref{chi}) constitute the basic theoretical framework for studying the cavity quantum electrodynamics in CEP cavity. In the following subsections, we derive the conditions of single-photon DBS in CEP cavity based on Eqs. (\ref{eq2})-(\ref{eq3}). 

\subsection{Vacancy-like dressed bound state}
The coupled cavity is the simplest model that supports the vacancy-like DBS, where the QE interacts with one of two cavities  \cite{vlbs}. At first glance our model is different from the coupled cavity proposed in Ref. \cite{vlbs}, it would be clear by changing the basis of cavity modes. To find its condition in CEP cavity, we rewrite $c_{ccw}$ and $c_{cw}$ in terms of the operators that represent the standing wave modes $c_1$ and $c_2$ \cite{pra2007}
\begin{equation}\label{eq4}
c_{c w}=\frac{1}{\sqrt{2}}\left(c_1+c_2\right), \quad c_{c c w}=\frac{1}{\sqrt{2}}\left(c_1-c_2\right)
\end{equation}
Substituting Eq. (\ref{eq4}) into Eq. (\ref{eq3}), we obtain $d\vec{s}/dt=-i \mathbf{M}_s \vec{s}$ with $\vec{s}=\left[\left\langle\sigma_{-}\right\rangle,\left\langle c_1\right\rangle,\left\langle c_2\right\rangle\right]^T$. The matrix $\mathbf{M}_s$ takes the form
\begin{equation}\label{eq5}
\mathbf{M}_s=\left[\begin{array}{ccc}
\omega_c & \sqrt{2} g & 0 \\
\sqrt{2} g & \omega_c-i \frac{\kappa\left(1+e^{i \phi}\right)}{2} & i \frac{\kappa}{2} e^{i \phi} \\
0 & -i \frac{\kappa}{2} e^{i \phi} & \omega_c-i \frac{\kappa\left(1-e^{i \phi}\right)}{2}
\end{array}\right]
\end{equation}

\begin{figure}[b]
\centering\includegraphics[width=0.95\linewidth]{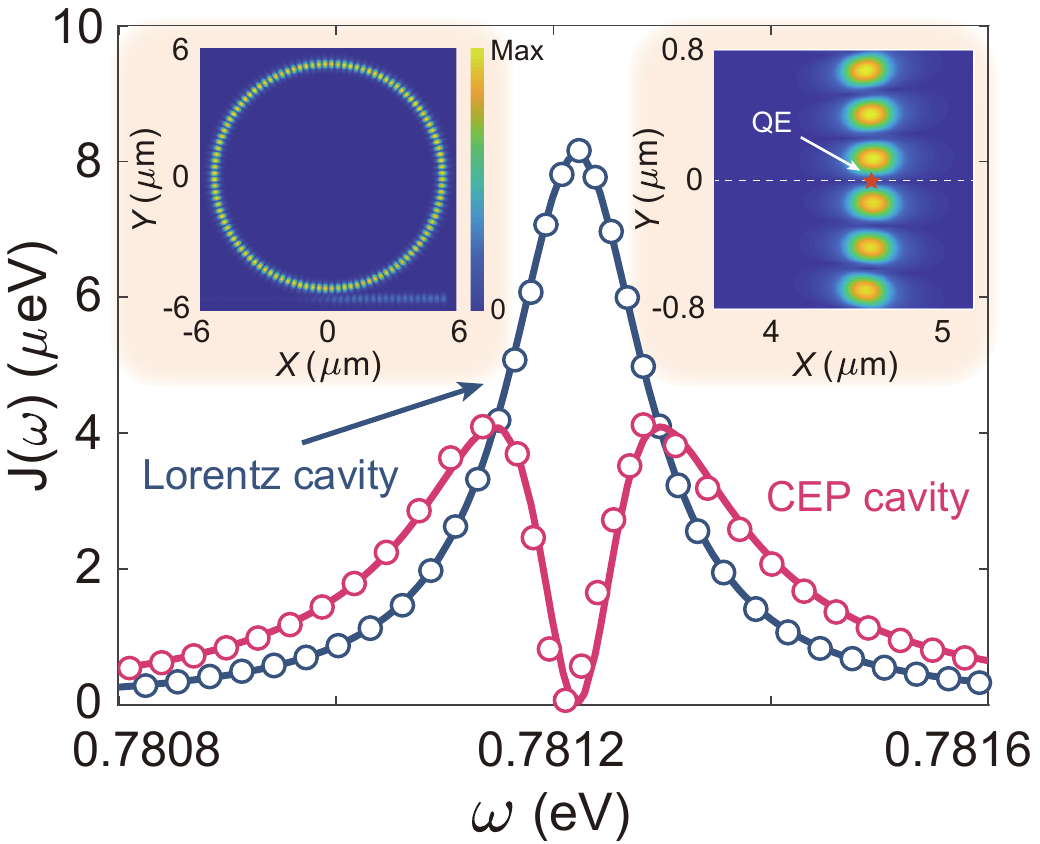}
\caption{Spectral density of a realistic CEP cavity with parameters: Outer radius $R=5{\mu}\mathrm{m}$, width $w=0.25{\mu}\mathrm{m}$, refractive index $n_c=3.47$, edge-to-edge separation to the waveguide $d=0.2{\mu}\mathrm{m}$. The width of waveguide is $d$ and the mirror is made of 100-nm thick silver. The refractive index of background medium is $n_b=1.44$. The blue circles plot the numerical result of Lorentz cavity (CEP cavity without the mirror at the end), while the blue solid line shows the fitting result with Lorentz spectral function. The pink solid line and circles represent the analytical and numerical results of CEP cavity, respectively. The insets show the electric field distribution of vacancy-like DBS. }
\label{fig2}
\end{figure}

\begin{figure*}[t]
\centering\includegraphics[width=0.95\linewidth]{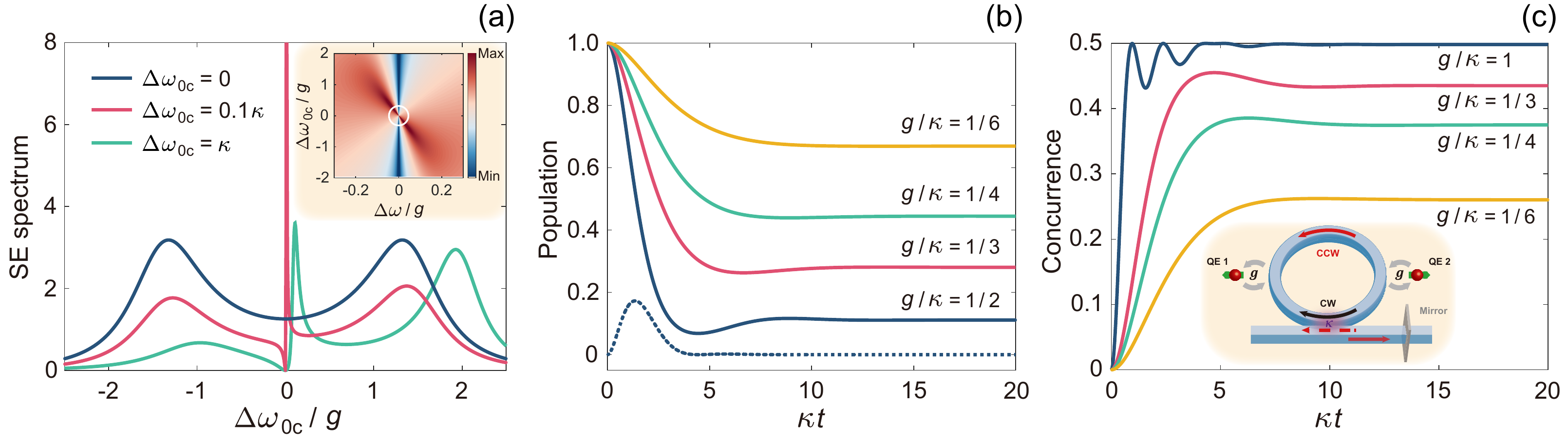}
\caption{(a) SE spectrum versus the QE-cavity detuning $\Delta\omega_{0c}=\omega_0-\omega_c$ for $g=\kappa$. The white circle in the inset indicates the vacancy-like DBS. (b) and (c) SE dynamics and dynamical concurrence of vacancy-like DBS with an excited QE for various $g/\kappa$, respectively. The blue dashed line in (b) shows the population of $c_1$ for $g/\kappa=1/2$. The inset in (c) illustrates the configuration of SEG.}
\label{fig3}
\end{figure*}

It shows that the QE is decoupled from the standing wave mode $c_2$. The vacancy-like DBS forms when the decay of $c_2$ is vanishing, i.e., $\phi=2n\pi$ ($n$ is an integer). In this case, the eigenstate is 
\begin{equation}\label{eq6}
\left|\psi_{V L}\right\rangle=\left(\frac{-i \kappa}{\sqrt{8 g^2+\kappa^2}}, 0, \frac{2 \sqrt{2} g}{\sqrt{8 g^2+\kappa^2}}\right)^T
\end{equation}
with energy $\omega_{VL}=\omega_c$, the same as bare QE. It indicates that the photon cannot be found at $c_1$ since its wavefunction is zero. As a consequence, the DBS can exist despite the presence of cavity dissipation, a feature not reported in the previous work \cite{vlbs}. Accordingly, the standing wave mode $c_1$ is called the vacancy cavity. Fig. \ref{fig1}(b) illustrates the concept of vacancy-like DBS in our model.

The existence of vacancy-like DBS can be confirmed by inspecting the spectral density of CEP cavity, which is given by $J(\omega)=\operatorname{Re}\int_{-\infty}^{+\infty} d \tau e^{i \omega \tau} 2 g^2\left\langle c_1^{\dagger}(\tau) c_1(0)\right\rangle$ for $\phi=2n\pi$ \cite{prl2018,prbfano}, where the two-time correlation $\left\langle c_1^{\dagger}(\tau) c_1(0)\right\rangle$ can be calculated in similar fashion as $\left\langle \sigma_{+}^{\dagger}(\tau) \sigma_{-}(0)\right\rangle$ using the quantum regression theorem. With the initial conditions $\left\langle c_1^{\dagger}(0) c_1(0)\right\rangle = 1$ and $\left\langle c_1^{\dagger}(0) c_2(0)\right\rangle = 0$, the spectral density can be analytically obtained  
\begin{equation}\label{eq8}
J(\omega)=\frac{2 g^2 \kappa}{\pi}\left[\frac{\omega-\omega_c}{\left(\omega-\omega_c\right)^2+\left(\frac{\kappa}{2}\right)^2}\right]^2
\end{equation}
It indicates that on resonance ($\omega=\omega_c$) the spectral density is zero, implying the null electric field amplitude at QE location. Physically, it means that there is no available channel for QE to decay, consistent with the nature of vacancy-like DBS. Fig. \ref{fig2} compares the analytical spectral density of a realistic CEP cavity (pink solid line) with the numerical results obtained from electromagnetic simulations (pink circles), where a good accordance can be seen. The insets of Fig. \ref{fig2} show the electric field distribution at $J(\omega)=0$, where we can see that the QE is located at a node of cavity modes, and thus decoupled from $c_1$, contract to the conventional Lorentz cavity (blue line and circles), i.e., CEP cavity without the mirror, where the QE location is exactly the antinode of standing wave mode. We thus understand that in CEP cavity, the physical origin of vacancy-like DBS can be interpreted as a result of the destructive interference between the cavity field of CCW mode and the reflected field of CW mode.

Fig. \ref{fig3}(a) shows that the SE spectrum is triplet deviated from DBS, with a Fano-type lineshape around the cavity resonance. As the QE energy approaches to the cavity resonance, the central peak in SE spectrum becomes sharper and goes upwards; on resonance ($\omega_0=\omega_c$) the central peak disappears, implying the formation of vacancy-like DBS. In this case, the SE spectrum exhibits a symmetrical Rabi splitting with a width of approximately $\sqrt{2} g$ (blue line).

Fig. \ref{fig3}(b) plots the time evolution of the population on the excited QE. It can be seen that the population of QE can be fractionally trapped for various $g/\kappa$. As the eigenstate $\left|\psi_{V L}\right\rangle$ indicates, the steady-state population remains finite but declines as $g$ increases due to the stronger population transfer from the QE to the cavity. By contrast, the population of $c_1$ is depleted at the steady state (blue dashed line) as expected. 

\begin{figure*}[t]
\centering\includegraphics[width=0.95\linewidth]{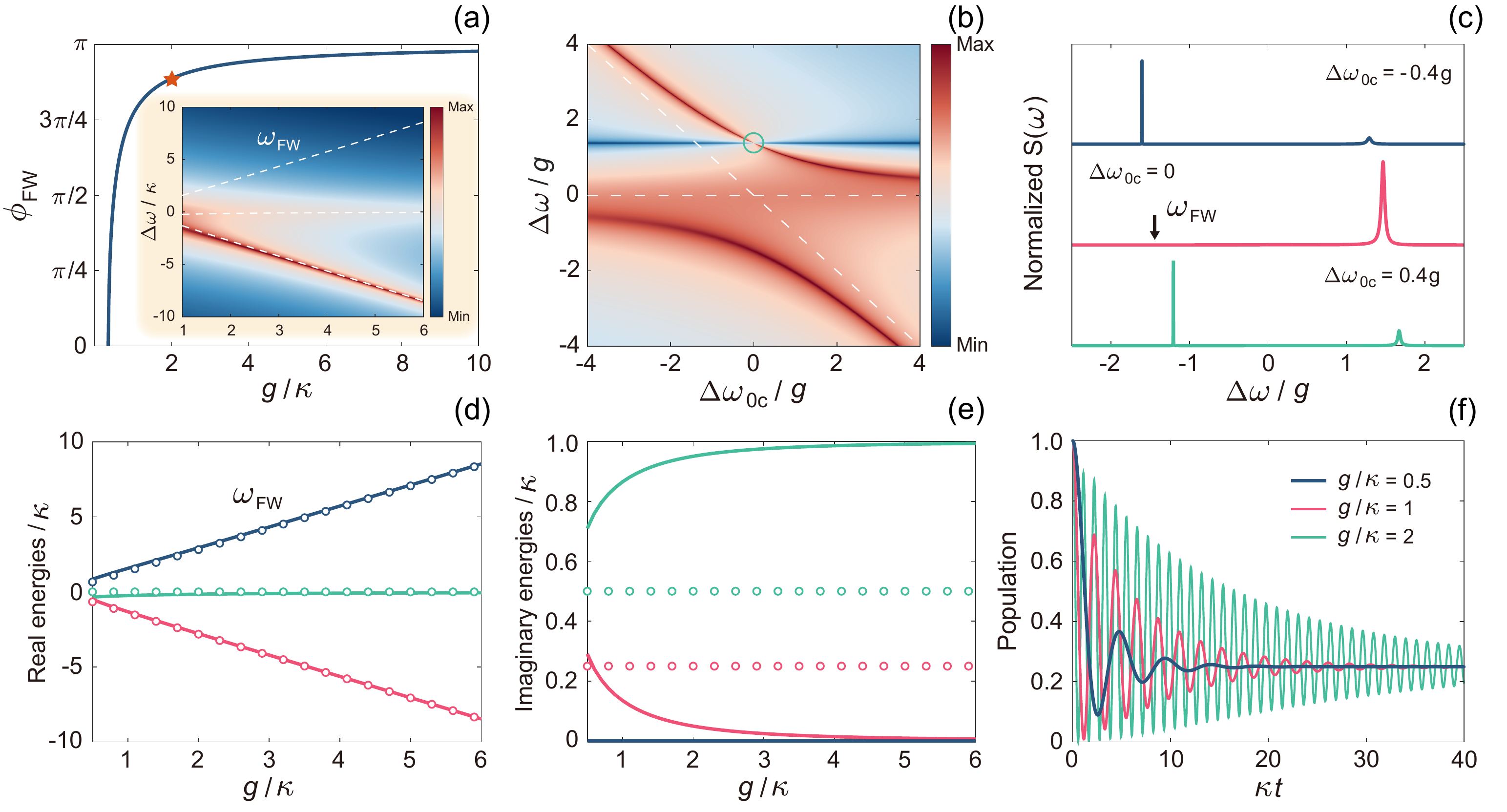}
\caption{(a) Condition of Friedrich-Wintgen DBS versus $g/\kappa$. The red star indicates the parameters for (b). The inset shows the corresponding SE spectrum, where the white dashed lines track the real eigenenergies. (b) Logarithmic plot of SE spectrum versus the QE-cavity detuning $\Delta\omega_{0c}$. SE spectra for $\Delta\omega_{0c}=-0.4g$, $0$, and $0.4g$ are shown in (c). The black arrow indicates the energy of vanishing Rabi peak. (d) and (e) The real and imaginary parts of eigenenergies versus $g/\kappa$, respectively, for CEP cavity with Friedrich-Wintgen DBS (solid lines) and Lorentz cavity (circles). Note that the blue circles are overlapped with the pink in (e). (f) SE dynamics with Friedrich-Wintgen DBS for various $g/\kappa$. }
\label{fig4}
\end{figure*}

Since the vacancy-like DBS occurs in the case of resonant QE-cavity coupling, it is beneficial for numerous quantum-optics applications especially for those involving the energy transfer mediated by cavity, such as the spontaneous entanglement generation (SEG) between qubits \cite{prl1993,prbce,pra2016}. It is straightforward to extend our model to multi-QE case by replacing $H$ in Eq. (\ref{eq1}) with the multi-QE Hamiltonian $H^M$, which is given by $H^M=H_0^M+H_I^M$, where $H_0^M=\omega_0 \sum_i \sigma_{+}^{(i)} \sigma_{-}^{(i)}+\omega_c c_{ccw}^{\dagger} c_{ccw}+\omega_c c_{cw}^{\dagger} c_{cw}$ and $H_I^M=g \sum_i\left( c_{ccw}^{\dagger} \sigma_{-}^{(i)}+ \sigma_{+}^{(i)} c_{ccw}\right)+g \sum_i\left( c_{cw}^{\dagger} \sigma_{-}^{(i)}+ \sigma_{+}^{(i)} c_{cw}\right)$. With an initially excited qubit, the generated entanglement between two qubits is quantified by the concurrence $C(t)=2\left|C_{e g}(t) C_{g e}^*(t)\right|$ \cite{prl1993,lywnp}, where $C_{eg}(t)$ and $C_{ge}(t)$ are the probability amplitudes of two single-excitation states that one qubit in the excited state while another in the ground state (detailed derivation is given in Appendix. \ref{ac}). The inset of Fig. \ref{fig3}(c) shows the illustration of SEG mediated by CEP cavity, where the long-distance entanglement can be generated between two qubits. As the results shown in Fig. \ref{fig3}(c), the higher and faster steady-state entanglement can be achieved as $g$ increases and reaches the maximum 0.5 with $g/\kappa=1$. Since the $Q$ factor of WGM cavity is typically $10^5$ at near infrared \cite{pnas}, the vacancy-like DBS allows for a fast and perfect entanglement without requiring a demanding coupling strength between the qubits and the cavity. In addition, we can see that Figs. \ref{fig3}(b) and (c) present opposite trends for a large $g$. It indicates that the strong population transfer from QE to cavity is unfavourable for population trapping of a single QE, but can lead to the efficient QE-QE interaction mediated by cavity, and thus is beneficial for achieving SEG with long-lived entanglement.

\subsection{Friedrich-Wintgen dressed bound state}
Different from vacancy-like DBS, Friedrich-Wintgen DBS originates from the destructive interference of two coupling pathways, one mediated by QE while another mediated by waveguide, as Fig. \ref{fig1}(c) depicts. To derive the condition of Friedrich-Wintgen DBS, we recast $\mathbf{M}_c$ in the following form \cite{opticabs}
\begin{equation}\label{eq9}
\mathbf{M}_c=H_B-i \Gamma
\end{equation}
with the Hermitian part giving rise to real energy for DBS
\begin{equation}\label{eq10}
H_B=\left[\begin{array}{ccc}
\omega_c & g & g \\
g & \omega_c & i \frac{\kappa}{2} e^{-i \phi} \\
g & -i \frac{\kappa}{2} e^{i \phi} & \omega_c
\end{array}\right]
\end{equation}
and the dissipative operator governing the imaginary part of eigenenergies
\begin{equation}\label{eq11}
\Gamma=D^{\dagger} D=\left[\begin{array}{ccc}
0 & 0 & 0 \\
0 & \frac{\kappa}{2} & \frac{\kappa}{2} e^{-i \phi} \\
0 & \frac{\kappa}{2} e^{i \phi} & \frac{\kappa}{2}
\end{array}\right]
\end{equation}

\noindent Subsequently, we can determine the coupling matrix $D=\left(0, \sqrt{\kappa / 2}, \sqrt{\kappa / 2} e^{-i \phi}\right)$ and introduce an unnormalized null vector of $D$, $\left|\psi_0\right\rangle=\left(\alpha,-e^{-i \phi}, 1\right)^T$, satisfying $D\left|\psi_0\right\rangle = 0$, where $\alpha$ is an undetermined coefficient. The Friedrich-Wintgen DBS appears when $\left|\psi_0\right\rangle$ fulfills $H_B\left|\psi_0\right\rangle=\omega_{F W}\left|\psi_0\right\rangle$. The solutions yield the energy and condition of Friedrich-Wintgen DBS
\begin{equation}\label{eq12}
\omega_{F W}=\omega_c \pm \frac{\sqrt{8 g^2-\kappa^2}}{2}
\end{equation}
\begin{equation}\label{eq13}
\phi_{F W}=-i \ln \left(-\frac{\left(4 g^2-\kappa^2\right) \pm i \kappa \sqrt{8 g^2-\kappa^2}}{4 g^2}\right)
\end{equation}

Fig. \ref{fig4}(a) plots $\phi_{F W}$ versus $g/\kappa$, where it shows that $\phi_{F W}$ tends to $\pi$ for a large $g$. With $\phi_{F W}$, there are only two peaks seen in SE spectrum, the Rabi peak corresponding to Friedrich-Wintgen DBS is invisible due to the vanishing linewidth, as the inset of Fig. \ref{fig4}(a) shows. On the other hand, by continuously varying the QE-cavity detuning, we observe an unusual behavior of strong-coupling anticrossing shown in Figs. \ref{fig4}(b) and (c), where the linewidth of one of bands is narrower and the peak disappears at a specific frequency (on resonance here, see the green circle in Fig. \ref{fig4}(b) and the pink line in \ref{fig4}(c)), a signature of Friedrich-Wintgen-type bound states \cite{prl2008,pra1985}. The real and imaginary parts of eigenenergies versus $g/\kappa$ are plotted in Figs. \ref{fig4}(d) and (e), respectively, where it shows that the real energies of CEP cavity are nearly the same as Lorentz cavity, while the imaginary parts are dissimilar. It is worth noting that the linewidth of the remaining Rabi peak significantly narrows for $g>\kappa$ compared to the Lorentz cavity, and approaches to zero as $g$ gradually increases, see the pink solid line shown in Fig. \ref{fig4}(e). The corresponding linewidth is found to be $\sim\left(1+\cos \left(\phi_{F W}\right)\right) / 2$ for $g \gg \kappa$. The linewidth narrowing of dressed states is accompanied by the decay suppression of Rabi oscillation in time domain, see the SE dynamics for various $g/\kappa$ shown in Fig. \ref{fig4}(f). Therefore, for Friedrich-Wintgen DBS, a large $g$ is beneficial to achieve a long decoherence time. 

\begin{figure}[b]
\centering\includegraphics[width=0.96\linewidth]{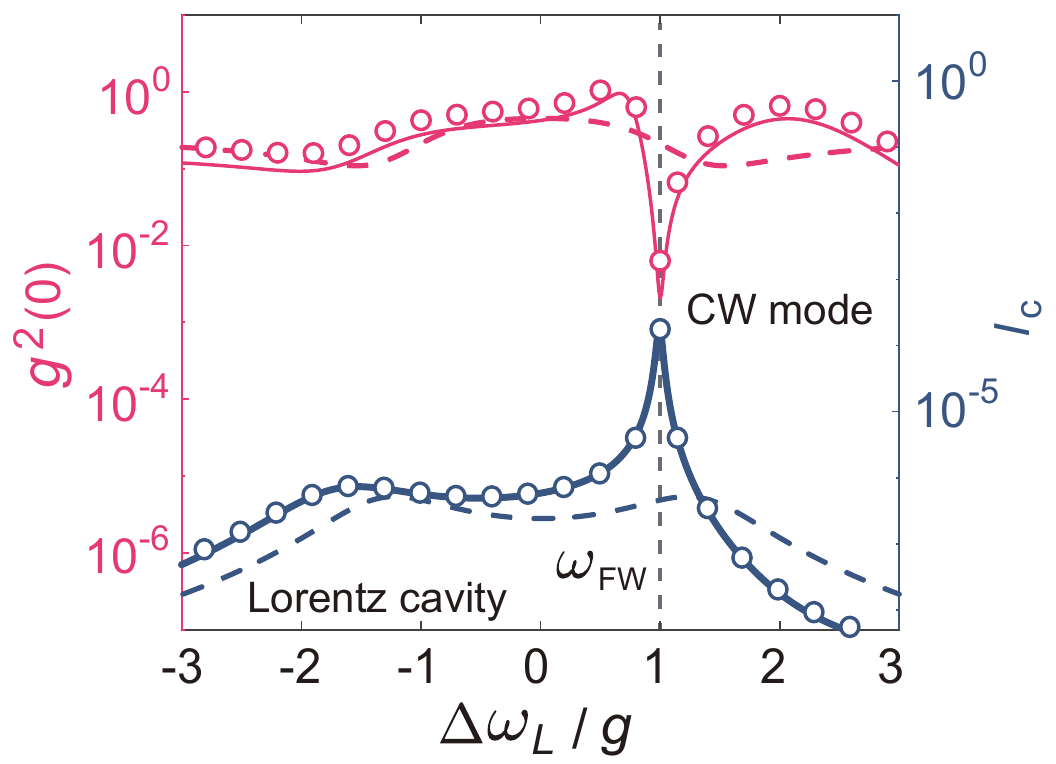}
\caption{Comparison of single-photon blockade of CEP cavity (circles for numerical results and solid lines for analytical results) with Lorentz cavity (dashed lines). The results are obtained by implementing a driving Hamiltonian $H_d=\Omega\left(e^{-i\omega_Lt}\sigma_{+}+e^{i\omega_Lt}\sigma_{-}\right)$ and a Liouvillian superoperator for QE dissipation $\gamma \mathcal{L}\left[\sigma_{-}\right] \rho$ in Eq. (\ref{eq1}). The numerical results are obtained using QuTip \cite{qutip}. The parameters used in the simulations are $g=\kappa/2$, $\gamma=\kappa/20$, and $\Omega=10^{-2} \gamma$. The analytical expressions of $I_c$ and $g^{(2)}(0)$ are derived in Appendix. \ref{ad}. The vertical dashed line indicates $\omega_{FW}$ given by Eq. (\ref{eq12}). }
\label{fig5}
\end{figure}

Though both are DBS in the same cavity QED system, there are two great differences between the vacancy-like DBS and the Friedrich-Wintgen DBS. One difference is that the steady-state population of the former depends on the coupling strength $g$ (see Fig. \ref{fig3}(b)) while the latter does not, as Fig. \ref{fig4}(f) shows. We find that half the energy can be trapped in the system via Friedrich-Wintgen DBS and the steady-state population of QE is 1/4 irrespective of $g$. Another difference lies in the energy of DBS. The energy of vacancy-like DBS is equal to bare QE for any $g$, while Friedrich-Wintgen DBS occurs at one of the anharmonic energy levels that the energy spacing is proportional to $g$. This feature offers Friedrich-Wintgen DBS unique potential for single-photon generation utilizing the photon blockade effect \cite{valle,xkyprap,lfl}. Fig. \ref{fig5} compares the performance of single-photon blockade of CEP cavity with that of Lorentz cavity. It shows that the best performance is achieved at Friedrich-Wintgen DBS (vertical dashed line), where both the single-photon efficiency $I_c=\left\langle c_{c w}^{\dagger} c_{c w}\right\rangle$ and the photon correlation $g^{(2)} (0) = \left\langle c_{c w}^{\dagger} c_{c w}^{\dagger} c_{c w} c_{c w}\right\rangle / I^2_c$ manifest a remarkable enhancement by over two orders of magnitude compared to the dressed states in conventional Lorentz cavity (dashed lines).

\section{Conclusion}
In conclusion, we demonstrate and unveil the origin of DBS in a prototypical microring resonator operating at CEP, which are classified into two types, the vacancy-like and Friedrich-Wintgen-type bound states. DBS studied in this work exists in the single-photon manifold, while the principles can be applied to higher-excitation manifold for exploring multi-photon DBS. Besides the SEG and single-photon generation demonstrated here, we envision the prominent advantages of DBS in diverse applications, such as quantum logic gate operation and quantum sensing, due to the long decoherence time and extremely sharp lineshape of DBS. We believe our work not only deepens the understanding of DBS at CEP, but also paves the way for harnessing the non-Hermitian physics to manipulate quantum states in a novel way.

\begin{acknowledgments}
Y. Lu acknowledges the support of the National Natural Science Foundation of China (Grant No. 62205061) and the Postdoctor Startup Project of Foshan (Grant No. BKS205043). Z. Liao is supported by the National Key R\&D Program of China (Grant No. 2021YFA1400800) and the Natural Science Foundations of Guangdong (Grant No. 2021A1515010039).
\end{acknowledgments}

\appendix

\section{Derivation of the extended cascaded quantum master equation}\label{aa}

The extended cascaded quantum master equation (QME) in Eq. (\ref{eq1}) can be derived from tracing out the waveguide modes based on the model depicted in Fig. \ref{fig1}(c). The system Hamiltonian including the waveguide modes is written as ($\hbar = 1$)
\begin{equation}
H_S=H+H_B+H_{S B}
\end{equation}
\begin{widetext}
\noindent where $H=H_0+H_I$ is given in Eq. (\ref{eq1}). $H_B$ is the free Hamiltonian of waveguide
\begin{equation}
H_B=\int d \omega \omega b_R^{\dagger} b_R
\end{equation}
and $H_{SB}$ describes the Hamiltonian of cavity-waveguide interaction
\begin{equation}
H_{S B}=i \sum_{j=c c w_{,} c w} \int d \omega \sqrt{\frac{\kappa}{2 \pi}} b_R^{\dagger} e^{-i k x_j} c_j+H . c .
\end{equation}
where $b_R$ is the bosonic annihilation operator of the right-propagating waveguide mode with frequency $\omega$ and wave vector $k=\omega_c/v$ with $v$ being the group velocity. $x_{ccw}$ and $x_{cw}$ are the locations of CCW mode and the mirrored CW mode. Applying the transformation $\widetilde{H}=U H U^{\dagger}-i dU/dt U^{\dagger}$ with $U = \exp \left[i\left(\omega_c \sum_{j=c c w, c w} c_j^{\dagger} c_j+\int d \omega \omega b_R^{\dagger} b_R\right)\right]$, we have
\begin{equation}
\widetilde{H}_{S B}(t)=i \sum_{j=c c w, c w} \int d \omega \sqrt{\frac{\kappa}{2 \pi}} b_R^{\dagger} e^{i\left(\omega-\omega_c\right) t} e^{-i \omega x_j / v} c_j+H . c .
\end{equation}
The equation of motion of $b_R$ can be obtained from the Heisenberg equation
\begin{equation}
\frac{d}{d t} b_R(t)=\sum_{j=c c w_{,} c w} \sqrt{\frac{\kappa}{2 \pi}} c_j e^{i\left(\omega-\omega_c\right) t} e^{-i \omega x_j / v}
\end{equation}
The above equation can be formally integrated to obtain
\begin{equation}
b_R(t)=\sum_{j=c c w, c w} \int_0^t d \tau \sqrt{\frac{\kappa}{2 \pi}} c_j e^{i\left(\omega-\omega_c\right) \tau} e^{-i \omega x_j / v}
\end{equation}
where we have taken $b_R (0)=0$ since the waveguide is initially in the vacuum state. On the other hand, the equation of motion of arbitrary operator $O$ is given by
\begin{equation}
\frac{d}{d t} O(t)=\sum_{j=c c w, c w} \int d \omega \sqrt{\frac{\kappa}{2 \pi}}\left\{b_R^{\dagger}(t) e^{i\left(\omega-\omega_c\right) t} e^{-i \omega x_j / v}\left[O(t), c_j(t)\right]-\left[O(t), c_j^{\dagger}(t)\right] b_R(t) e^{-i\left(\omega-\omega_c\right) t} e^{i \omega x_j / v}\right\}
\end{equation}
Substituting $b_R (t)$ into the above equation, we have
\begin{equation}\label{a8}
\begin{aligned}
\frac{d}{d t} O(t)=\frac{\kappa}{2 \pi} \sum_{j, l=c c w, c w}&  \int_0^t d \tau \int d \omega\left\{e^{i\left(\omega-\omega_c\right)(t-\tau)} e^{-i \omega x_j / v} c_l^{\dagger}(\tau)\left[O(t), c_j(t)\right]\right. \\
& \left.-\left[O(t), c_j^{\dagger}(t)\right] c_l(\tau) e^{-i\left(\omega-\omega_c\right)(t-\tau)} e^{i \omega x_{j l} / v}\right\} \\
&
\end{aligned}
\end{equation}
where $x_{jl}=x_j-x_l$. We apply the Markov approximation by assuming the time delay $x_{jl}/v$ between the CCW mode and the mirrored CW mode can be neglected. Therefore,
\begin{equation}\label{a9}
\begin{gathered}
\frac{\kappa}{2 \pi} \sum_{l=c c w, c w} \int_0^t d \tau \int d \omega e^{i\left(\omega-\omega_c\right)(t-\tau)} e^{-i \omega x_{j l} / v} c_l^{\dagger}(\tau)=\kappa \sum_{l=c c w, c w} \int_0^t d \tau \delta\left(t-\frac{x_{j l}}{v}-\tau\right) e^{-i k x_{j l}} c_l^{\dagger}(\tau) \\
\approx \frac{\kappa}{2} c_j^{\dagger}(t)+\kappa \sum_{l=c c w, c w} \Theta\left(t-\frac{x_{j l}}{v}\right) e^{-i k x_{j l}} c_l^{\dagger}(t)
\end{gathered}
\end{equation}
where $x_{jl}>0$ and $\Theta(t)$ is the step function. With Eq. (\ref{a9}) and taking the averages of Eq. (\ref{a8}), we have
\begin{equation}
\begin{aligned}
\frac{d}{d t}\langle O(t)\rangle=\frac{\kappa}{2} & \sum_{j=c c w, c w}\left\{\left\langle c_j^{\dagger}(t)\left[O(t), c_j(t)\right]\right\rangle-\left\langle\left[O(t), c_j^{\dagger}(t)\right] c_j(t)\right\rangle\right\} \\
& +\kappa \sum_{j, l=c c w, c w, j \neq l}\left\{e^{-i k x_{j l}}\left\langle c_l^{\dagger}(t)\left[O(t), c_j(t)\right]\right\rangle-e^{i k x_{j l}}\left\langle\left[O(t), c_j^{\dagger}(t)\right] c_l(t)\right\rangle\right\}
\end{aligned}
\end{equation}
Since $\langle O(t)\rangle=\operatorname{Tr}[O(t) \rho(0)]=\operatorname{Tr}[O \rho(t)]$, we can simplify the averages of operators in the above equation by using the cyclic property of trace. For example,
\begin{equation}
\left\langle\left[O(t), c_j^{\dagger}(t)\right] c_j(t)\right\rangle=\operatorname{Tr}\left[O c_j^{\dagger} c_j \rho(t)-c_j^{\dagger} O c_j \rho(t)\right]=\operatorname{Tr}\left[O c_j^{\dagger} c_j \rho(t)-O c_j \rho(t) c_j^{\dagger}\right]=\operatorname{Tr}\left\{O\left[c_j^{\dagger}, c_j \rho(t)\right]\right\}
\end{equation}
Therefore, we can obtain a QME in the following form 
\begin{equation}
\begin{gathered}
\frac{d}{d t} \rho(t)=-i[H, \rho(t)]+\frac{\kappa}{2} \sum_{j=c c w, c w}\left\{\left[c_j, \rho(t) c_j^{\dagger}\right]-\left[c_j^{\dagger}, c_j \rho(t)\right]\right\} \\
+\kappa \sum_{j, l=c c w, c w, j \neq l}\left\{e^{-i k x_{j l}}\left[c_j, \rho(t) c_l^{\dagger}\right]-e^{i k x_{j l}[}\left[c_j^{\dagger}, c_l \rho(t)\right]\right\}
\end{gathered}
\end{equation}
Note that $kx_{jl}=\phi$, and thus $j=cw$ and $l=ccw$ in the third term on the right-hand side. In addition, the second term on the right-hand side can be expanded and rewritten using the Liouvillian superoperator. We thus arrive at the extended cascaded QME in Eq. (\ref{eq1}).

\section{Derivation of the spontaneous emission spectrum}\label{ab}

The spontaneous emission (SE) spectrum, also called the polarization spectrum, reflects the local dynamics of a quantum emitter (QE). The SE spectrum is given by $S(\omega)=\lim _{t \rightarrow \infty} 2 \operatorname{Re}\left[\int_0^{\infty} d \tau\left\langle\sigma_{+}(t+\tau) \sigma_{-}(t)\right\rangle e^{i \omega \tau}\right]$, where the correlation $\left\langle\sigma_{+}(t+\tau) \sigma_{-}(t)\right\rangle$ can be solved from Eqs. (\ref{eq2})-(\ref{eq3}) using the quantum regression theorem, which yields the following equations of motion
\begin{equation}
\frac{d}{d \tau}\left[\begin{array}{c}
\left\langle\sigma_{+}(\tau) \sigma_{-}(0)\right\rangle \\
\left\langle\sigma_{+}(\tau) c_{c c w}(0)\right\rangle \\
\left\langle\sigma_{+}(\tau) c_{c w}(0)\right\rangle
\end{array}\right]=\left[\begin{array}{ccc}
\omega_0 & g & g \\
g & \omega_c-i \frac{\kappa}{2} & 0 \\
g & -i \kappa e^{i \phi} & \omega_c-i \frac{\kappa}{2}
\end{array}\right]\left[\begin{array}{c}
\left\langle\sigma_{+}(\tau) \sigma_{-}(0)\right\rangle \\
\left\langle\sigma_{+}(\tau) c_{c c w}(0)\right\rangle \\
\left\langle\sigma_{+}(\tau) c_{c w}(0)\right\rangle
\end{array}\right]
\end{equation}
Using the initial conditions $\left\langle\sigma_{+}(0) \sigma_{-}(0)\right\rangle=1$, $\left\langle\sigma_{+}(0) c_{ccw}(0)\right\rangle=0$, and $\left\langle\sigma_{+}(0) c_{cw}(0)\right\rangle=0$, the above correlations can be easily obtained by taking the Laplace transform $\langle O(\tau)\rangle \rightarrow\langle O(s)\rangle$ 
\begin{equation}
s\left[\begin{array}{c}
\left\langle\sigma_{+} \sigma_{-}(s)\right\rangle \\
\left\langle\sigma_{+} c_{c c w}(s)\right\rangle \\
\left\langle\sigma_{+} c_{c w}(s)\right\rangle
\end{array}\right]=\left[\begin{array}{ccc}
\omega_0 & g & g \\
g & \omega_c-i \frac{\kappa}{2} & 0 \\
g & -i \kappa e^{i \phi} & \omega_c-i \frac{\kappa}{2}
\end{array}\right]\left[\begin{array}{c}
\left\langle\sigma_{+} \sigma_{-}(s)\right\rangle \\
\left\langle\sigma_{+} c_{c c w}(s)\right\rangle \\
\left\langle\sigma_{+} c_{c w}(s)\right\rangle
\end{array}\right]+\left[\begin{array}{l}
1 \\
0 \\
0
\end{array}\right]
\end{equation}
The solutions are given by
\begin{equation}
\left\langle\sigma_{+} \sigma_{-}(s)\right\rangle=\frac{1}{s+i \omega_0+\frac{g^2}{s+i\left(\omega_c-i \frac{\kappa}{2}\right)}\left[2-\frac{\kappa e^{i \phi}}{s+i\left(\omega_c-i \frac{\kappa}{2}\right)}\right]}
\end{equation}
Transforming into the frequency domain by replacing $s=-i\omega$, we have
\begin{equation}
\left(-i\left(\omega-\omega_0\right)+\frac{2 g^2}{-i\left(\omega-\omega_c\right)+\kappa+i \frac{\left(\frac{\kappa}{2}\right)^2}{\omega-\omega_c}}\right)\left\langle\sigma_{+} \sigma_{-}(\omega)\right\rangle=1
\end{equation}
Therefore, 
\begin{equation}\label{b5}
\left\langle\sigma_{+} \sigma_{-}(\omega)\right\rangle=\frac{i}{\left(\omega-\omega_0\right)-g^2\left\{\frac{2}{\left(\omega-\omega_c\right)+i \frac{k}{2}}-\frac{i \kappa e^{i \phi}}{\left[\left(\omega-\omega_c\right)+i \frac{\kappa}{2}\right]^2}\right\}}
\end{equation}
We identify the response function of CEP cavity as 
\begin{equation}
\chi(\omega)=\frac{2}{\left(\omega-\omega_c\right)+i \frac{\kappa}{2}}-\frac{i \kappa e^{i \phi}}{\left[\left(\omega-\omega_c\right)+i \frac{\kappa}{2}\right]^2}
\end{equation}
where the first term of right-hand side denotes the usual Lorentz response, with a factor 2 representing the coupling of QE to two cavity modes. The second term of right-hand side demonstrates the characteristic of squared Lorentz response and thus is contributed by CEP. Eq. (\ref{b5}) can be rewritten as
\begin{equation}
\left\langle\sigma_{+} \sigma_{-}(\omega)\right\rangle=\frac{i}{\omega-\omega_0-\Delta(\omega)+i \frac{\Gamma(\omega)}{2}}
\end{equation}
Therefore, the SE spectrum is expressed as
\begin{equation}
S(\omega)=\frac{2}{\pi}\operatorname{Re}[\left\langle\sigma_{+} \sigma_{-}(\omega)\right\rangle]=\frac{1}{\pi} \frac{\Gamma(\omega)}{\left[\omega-\omega_0-\Delta(\omega)\right]^2+\left[\frac{\Gamma(\omega)}{2}\right]^2}
\end{equation}
with the photon induced Lamb shift
\begin{equation}
\Delta(\omega)=g^2 \operatorname{Re}[\chi(\omega)]=\frac{\left[\left(\omega-\omega_c\right)^2-\left(\frac{\kappa}{2}\right)^2\right]\left[2\left(\omega-\omega_c\right)+\kappa \sin (\phi)\right]+\kappa^2\left(\omega-\omega_c\right)[1-\cos (\phi)]}{\left[\left(\omega-\omega_c\right)^2+\left(\frac{\kappa}{2}\right)^2\right]^2}
\end{equation}
and the local coupling strength
\begin{equation}
\Gamma(\omega)=-2g^2 \operatorname{Im}[\chi(\omega)]=-2\frac{\left[\left(\omega-\omega_c\right)^2-\left(\frac{\kappa}{2}\right)^2\right] \kappa[1-\cos (\phi)]-\kappa\left(\omega-\omega_c\right)\left[2\left(\omega-\omega_c\right)+\kappa \sin (\phi)\right]}{\left[\left(\omega-\omega_c\right)^2+\left(\frac{\kappa}{2}\right)^2\right]^2}
\end{equation}
For vacancy-like bound state ($\phi=2n\pi$), the local coupling strength is 
\begin{equation}
\Gamma(\omega)=4 g^2 \kappa\left[\frac{\omega-\omega_c}{\left(\omega-\omega_c\right)^2+\left(\frac{\kappa}{2}\right)^2}\right]^2=2 \pi J(\omega)
\end{equation}
where $J(\omega)$ is given in Eq. (\ref{eq8}).

\section{Spontaneous entanglement generation at vacancy-like bound state}\label{ac}
The system Hamiltonian for spontaneous entanglement generation (SEG) is written as
\begin{equation}
H^M=H_0^M+H_I^M
\end{equation}
where $H_0^M$ and $H_I^M$ are given by
\begin{equation}
H_0^M=\omega_c \sum_{j=1,2} \sigma_{+}^{(j)} \sigma_{-}^{(j)}+\omega_c c_{c c w}^{\dagger} c_{c c w}+\omega_c c_{c w}^{\dagger} c_{c w}
\end{equation}
\begin{equation}
H_I^M=\sum_{j=1,2} g\left(\sigma_{-}^{(j)} c_{c c w}^{\dagger}+c_{c c w} \sigma_{+}^{(j)}\right)+g\left(\sigma_{-}^{(j)} c_{c w}^{\dagger}+c_{c w} \sigma_{+}^{(j)}\right)
\end{equation}
With the extended cascaded QME (Eq. (\ref{eq1})), we can obtain the effective Hamiltonian in the single-excitation subspace
\begin{equation}
\begin{aligned}
H_{\mathrm{eff}}=\omega_c \sum_{j=1,2} & \sigma_{+}^{(j)} \sigma_{-}^{(j)}+\left(\omega_c-i \frac{\kappa}{2}\right) c_{c c w}^{\dagger} c_{c c w}+\left(\omega_c-i \frac{\kappa}{2}\right) c_{c w}^{\dagger} c_{c w} \\
& +\sum_{j=1,2} g\left(\sigma_{-}^{(j)} c_{c c w}^{\dagger}+c_{c c w} \sigma_{+}^{(j)}\right)+g\left(\sigma_{-}^{(j)} c_{c w}^{\dagger}+c_{c w} \sigma_{+}^{(j)}\right)-i \kappa e^{i \phi} c_{c c w} c_{c w}^{\dagger}
\end{aligned}
\end{equation}
The corresponding state vector is given by
\begin{equation}
|\Psi(t)\rangle=C_{g g}(t)|g g 00\rangle+C_{e g}(t)|e g 00\rangle+C_{g e}(t)|g e 00\rangle+C_{10}(t)|g g 10\rangle+C_{01}(t)|g g 01\rangle
\end{equation}
where $\left|n_1 n_2 m p\right\rangle=\left|n_1\right\rangle \otimes\left|n_2\right\rangle \otimes|m\rangle \otimes|p\rangle$ with $\left|n_1\right\rangle$ and $\left|n_2\right\rangle$ representing that the QE is either in the excited state ($\left|n_1\right\rangle,\left|n_2\right\rangle=|e\rangle$) or in the ground state ($\left|n_1\right\rangle,\left|n_2\right\rangle=|g\rangle$), and $|m\rangle$ and $|p\rangle$ denoting that there is $m$ photon in the CCW mode and $p$ photon in the mirrored CW mode, respectively. With the Schrödinger equation $i d|\Psi(t)\rangle / d t=H_{\text {eff }}|\Psi(t)\rangle$, we can obtain the equations of coefficients
\begin{equation}
i \frac{d}{d t} C_{e g}(t)=\omega_c C_{e g}(t)+g C_{10}(t)+g C_{01}(t)
\end{equation}
\begin{equation}
i \frac{d}{d t} C_{g e}(t)=\omega_c C_{g e}(t)+g C_{10}(t)+g C_{01}(t)
\end{equation}
\begin{equation}
i \frac{d}{d t} C_{10}(t)=\left(\omega_c-i \frac{\kappa}{2}\right) C_{10}(t)+g C_{e g}(t)+g C_{g e}(t)
\end{equation}
\begin{equation}
i \frac{d}{d t} C_{01}(t)=\left(\omega_c-i \frac{\kappa}{2}\right) C_{01}(t)+g C_{e g}(t)+g C_{g e}(t)-i \kappa e^{i \phi} C_{10}(t)
\end{equation}
For vacancy-like bound state ($\phi=2n\pi$), the equations can be easily solved through the Laplace transform
\begin{equation}
C_{e g}(t)=\frac{8 g^2+\kappa^2+2 g e^{-\frac{\kappa}{2} t}[4 g \cos (2 g t)+\kappa \sin (2 g t)]}{16 g^2+\kappa^2}
\end{equation}
\begin{equation}
C_{g e}(t)=\frac{2 g e^{-\frac{\kappa}{2} t}\left[-4 g e^{\frac{\kappa}{2} t}+4 g \cos (2 g t)+\kappa \sin (2 g t)\right]}{16 g^2+\kappa^2}
\end{equation}
Then the dynamical concurrence can be obtained as $C(t)=2\left|C_{e g}(t) C_{g e}^*(t)\right|$.

\section{Single-photon generation at Friedrich-Wintgen bound state}\label{ad}
The single-photon generation through photon blockade requires the weak coherent pumping. In this section, we present a derivation of the analytical expressions for averaged photon number and zero-time-delay second-order correlation function of CW mode using the perturbation theory. A driving Hamiltonian is implemented in the extended cascaded QME for QE driven case, which is
\begin{equation}
H_{\text {driving }}=\Omega\left(e^{-i \omega_L t} \sigma_{+}+\sigma_{-} e^{i \omega_L t}\right)
\end{equation}
where $\omega_L$ is the frequency of laser field and $\Omega$ is the driving strength. Applying the unitary transformation $U=\exp \left[-i \omega_L\left(c_{c c w}^{\dagger} c_{c c w}+c_{c w}^{\dagger} c_{c w}+\sigma_{+} \sigma_{-}\right) t\right]$, we can obtain the effective Hamiltonian
\begin{equation}\label{d2}
H_{\mathrm{eff}}^t=H^t+E V
\end{equation}
with
\begin{equation}\label{d3}
\begin{aligned}
H^t=\Delta_0 & \sigma_{+} \sigma_{-}+\Delta_c c_{c c w}^{\dagger} c_{c c w}+\Delta_c c_{c w}^{\dagger} c_{c w}+g\left(\sigma_{-} c_{c c w}^{\dagger}+c_{c c w} \sigma_{+}\right)+g\left(\sigma_{-} c_{c w}^{\dagger}+c_{c w} \sigma_{+}\right) \\
& -i \kappa e^{i \phi} c_{c c w} c_{c w}^{\dagger}
\end{aligned}
\end{equation}
and
\begin{equation}
V=\Omega\left(\sigma_{+}+\sigma_{-}\right)
\end{equation}
where $\Delta_0=\Delta_{c L}-i \gamma / 2$ and $\Delta_c=\Delta_{c L}-i \kappa / 2$ with $\Delta_{c L}=\omega_c-\omega_L$ being the frequency detuning between the system and the laser field. $E$ is a perturbative parameter of laser intensity. Since the evaluation of $g^{(2)}(0)=\left\langle c_{c w}^{\dagger} c_{c w}^{\dagger} c_{c w} c_{c w}\right\rangle / I_c^2$ requires calculating the second-order correlation function of cavity operator, we expand the time-dependent wave function $|\Psi(t)\rangle$ in terms of $E$ as $|\Psi(t)\rangle=\sum_{l=2} E^l\left|\psi_l(t)\right\rangle$, where we have truncated the state space by two-excitation manifold and as a result, $\left|\psi_l(t)\right\rangle$ is expressed as
\begin{equation}
\left|\psi_l(t)\right\rangle=\sum_{n+m+p \leq 2, n=0,1} C_{n m p}^l|n\rangle_e|m\rangle_{c c w}|p\rangle_{c w}
\end{equation}
where $C_{nmp}^l$ is the coefficient of quantum state $|n\rangle_e|m\rangle_{c c w}|p\rangle_{c w}$ in $l$-order expansion, where there are $m$ photons in CCW mode and $p$ photons in CW mode, while the QE is either excited ($n=1$) or unexcited ($n=0$). For $l=1$ and $2$, the state vector is given by 
\begin{equation}
\left|\psi_1(t)\right\rangle=C_{100}^1|100\rangle+C_{010}^1|010\rangle+C_{001}^1|001\rangle
\end{equation}
\begin{equation}
\left|\psi_2(t)\right\rangle=C_{011}^2|011\rangle+C_{10}^2|110\rangle+C_{101}^2|101\rangle+C_{020}^2|020\rangle+C_{002}^2|002\rangle
\end{equation}
From the Schr{\"o}dinger equation $i d|\Psi(t)\rangle / d t=H_{\mathrm{eff}}^t|\Psi(t)\rangle$, we have
\begin{equation}\label{d9}
i \frac{d}{d t}\left|\psi_0(t)\right\rangle=H^t\left|\psi_0(t)\right\rangle
\end{equation}
\begin{equation}\label{d10}
i \frac{d}{d t}\left|\psi_l(t)\right\rangle=H^t\left|\psi_l(t)\right\rangle+V\left|\psi_{l-1}(t)\right\rangle
\end{equation}
Substituting $H^t$ (Eq. (\ref{d2})) and $V$ (Eq. (\ref{d3})) into Eqs. (\ref{d9}) and (\ref{d10}), we can obtain the following equations of motion for coefficients
\begin{equation}
i \frac{d}{d t} C_{100}^1=\Delta_0 C_{100}^1+g C_{010}^1+g C_{001}^1+\Omega
\end{equation}
\begin{equation}
i \frac{d}{d t} C_{010}^1=\Delta_c C_{010}^1+g C_{100}^1
\end{equation}
\begin{equation}
i \frac{d}{d t} C_{001}^1=\Delta_c C_{001}^1+g C_{100}^1-i \kappa e^{i \phi} C_{010}^1
\end{equation}
and $C_{000}^0 \approx 1$ due to the assumption of weak pump. The above equations yield
\begin{equation}
C_{001}^1=\Omega g \frac{\Delta_c+i \kappa e^{i \phi}}{D_1}
\end{equation}
with
\begin{equation}
D_1=\left|\begin{array}{ccc}
\Delta_0 & g & g \\
g & \Delta_c & 0 \\
g & -i \kappa e^{i \phi} & \Delta_c
\end{array}\right|
\end{equation}
Therefore, the averaged photon number of CW mode is given by
\begin{equation}
I_c=\left\langle\Psi(0)\left|c_{c w}^{\dagger} c_{c w}\right| \Psi(0)\right\rangle \approx\left|C_{001}^1\right|^2=\left|\Omega g \frac{\Delta_c+i \kappa e^{i \phi}}{D_1}\right|^2
\end{equation}
We can see that the eigenvalues of $D_1$ are the same as the matrix $\mathbf{M}_c$ in Eq. (\ref{eq9}), and thus the cavity photon $I_c$ diverges at the Friedrich-Wintgen DBS due to the zero decay, and the perfect single-photon purity can be achieved since $g^{(2)}(0)\propto I^{-2}_c$. This unphysical result comes from the truncation of state space with at most one excitation. $I_c$ will remain finite when taking into account the higher-order manifold. However, the analytical expression of $I_c$ predicts that the formation of bound state in the single-excitation subspace can produce prominent enhancement of both the efficiency and single-photon purity of single-photon blockade.

\noindent From Eq. (\ref{d10}), we can also obtain the equations of two-excitation subspace
\begin{equation}
i \frac{d}{d t} C_{011}^2=2 \Delta_c C_{011}^2+g C_{101}^2+g C_{110}^2-i \sqrt{2} \kappa e^{i \phi} C_{020}^2
\end{equation}
\begin{equation}
i \frac{d}{d t} C_{110}^2=\left(\Delta_0+\Delta_c\right) C_{110}^2+\sqrt{2} g C_{020}^2+g C_{011}^2+\Omega C_{010}^1
\end{equation}
\begin{equation}
i \frac{d}{d t} C_{101}^2=\left(\Delta_0+\Delta_c\right) C_{101}^2+g C_{011}^2+\sqrt{2} g C_{002}^2-i k e^{i \phi} C_{110}^2+\Omega C_{001}^1
\end{equation}
\begin{equation}
i \frac{d}{d t} C_{020}^2=2 \Delta_c C_{020}^2+\sqrt{2} g C_{110}^2
\end{equation}
\begin{equation}
i \frac{d}{d t} C_{002}^2=2 \Delta_c C_{002}^2+\sqrt{2} g C_{101}^2-i \sqrt{2} \kappa e^{i \phi} C_{011}^2
\end{equation}
We thus can obtain
\begin{equation}
\begin{aligned}
C_{002}^2=2 \sqrt{2} g D_2^{-1} & \left\{C_{001}^1\left\{\Delta_c\left[2 \Delta_c\left(\Delta_c+\Delta_0\right)-3 g^2\right]+i \kappa e^{i \phi}\left[\Delta_c\left(\Delta_c+\Delta_0\right)-2 g^2\right]\right\}\right. \\
& \left.+C_{010}^1\left\{\Delta_c g^2+i \kappa e^{i \phi}\left[\Delta_c\left(3 \Delta_c+\Delta_0\right)+g^2\right]-\kappa^2 e^{2 i \phi}\left(2 \Delta_c+\Delta_0\right)\right\}\right\}
\end{aligned}
\end{equation}
with
\begin{equation}
D_2=\left|\begin{array}{ccccc}
2 \Delta_c & g & g & -i \sqrt{2} \kappa e^{i \phi} & 0 \\
g & \Delta_0+\Delta_c & 0 & \sqrt{2} g & 0 \\
g & -i \kappa e^{i \phi} & \Delta_0+\Delta_c & 0 & \sqrt{2} g \\
0 & \sqrt{2} g & 0 & 2 \Delta_c & 0 \\
-i \sqrt{2} \kappa e^{i \phi} & 0 & \sqrt{2} g & 0 & 2 \Delta_c
\end{array}\right|=4 D_1\left[-2 g^2+\Delta_c\left(3 \Delta_c+2 \Delta_0\right)\right]+4 \Delta_c^4\left(2 \Delta_c+\Delta_0\right)
\end{equation}
Then the zero-time-delayed second-order correlation function is evaluated as
\begin{equation}
g^{(2)}(0)=\left\langle\Psi(0)\left|c_{c w}^{\dagger} c_{c w}^{\dagger} c_{c w} c_{c w}\right| \Psi(0)\right\rangle / I_c^2 \approx\left|C_{002}^2\right|^2 / I_c^2
\end{equation}

\end{widetext}

\nocite{*}

\bibliography{BS}

\end{document}